\documentclass{article}

\usepackage{arxiv}
\usepackage[utf8]{inputenc} % allow utf-8 input
\usepackage[T1]{fontenc}    % use 8-bit T1 fonts
\usepackage{hyperref}       % hyperlinks
\usepackage{url}            % simple URL typesetting
\usepackage{booktabs}       % professional-quality tables
\usepackage{amsfonts}       % blackboard math symbols
\usepackage{nicefrac}       % compact symbols for 1/2, etc.
\usepackage{microtype}      % microtypography
\usepackage{lipsum}         % Can be removed after putting your text content
\usepackage{graphicx}
\usepackage[numbers,sort&compress]{natbib}
\usepackage{doi}
\usepackage{amsmath}
\usepackage[inkscapeformat=png]{svg}
\usepackage{subcaption}  % For subfigures
\usepackage{array}
\usepackage{adjustbox}  % Add this to your preamble

\title{Deep Learning vs. Black-Scholes: Option Pricing Performance on Brazilian Petrobras Stocks}

\date{February 27, 2025}	% Here you can change the date presented in the paper title
\author{
  Joao Felipe Gueiros \quad 
  Hemanth Chandravamsi \quad
  Steven H. Frankel
  \vspace{5pt} \\
  Faculty of Mechanical Engineering, Technion - Israel Institute of Technology, Haifa, Israel 3200003 \\
  \texttt{\{joao.g, hemanth\}@campus.technion.ac.il}, \texttt{frankel@me.technion.ac.il}
}

\renewcommand{\shorttitle}{\textit{Deep Learning vs. Black-Scholes: Option Pricing Performance on Brazilian Petrobras Stocks}}

\hypersetup{
pdftitle={Deep Learning vs. Black-Scholes},
pdfsubject={q-bio.NC, q-bio.QM},
pdfauthor={Joao Gueiros},
pdfkeywords={First keyword, Second keyword, More},
}

\begin{document}
\maketitle

\begin{abstract}
This paper explores the use of deep residual networks for pricing European options on Petrobras, one of the world's largest oil and gas producers, and compares its performance with the Black-Scholes (BS) model. Using eight years of historical data from B3 (Brazilian Stock Exchange) collected via web scraping, a deep learning model was trained using a custom built hybrid loss function that incorporates market data and analytical pricing. The data for training and testing were drawn between the period spanning November 2016 to January 2025, using an 80-20 train-test split. The test set consisted of data from the final three months: November, December, and January 2025. The deep residual network model achieved a 64.3\% reduction in the mean absolute error for the 3-19 BRL (Brazilian Real) range when compared to the Black-Scholes model on the test set. Furthermore, unlike the Black-Scholes solution, which tends to decrease its accuracy for longer periods of time, the deep learning model performed accurately for longer expiration periods. These findings highlight the potential of deep learning in financial modeling, with future work focusing on specialized models for different price ranges.
\end{abstract}

% keywords can be removed

\section{Introduction}
The attempt to model and predict the stock market is not new. Even Isaac Newton, one of history’s greatest minds, famously failed in his investments, losing around 20,000 pounds at the time. Today, quantitative analysts at major hedge funds continue this pursuit, leveraging advanced mathematical modeling and machine learning to anticipate market movements and gain an edge over competitors.
A pivotal moment in the mathematical study of finance came with Louis Bachelier’s doctoral thesis \cite{Bachelier1900}, which introduced the first formal model of Brownian motion and its application to valuing stock options. His work laid the foundation for modern financial mathematics and influenced the development of widely used models, including the Black-Scholes model, which is essential for pricing options, and led Researchers Fischer Black and Myron Scholes to win the Nobel prize for Economy in 1997 \cite{black1973pricing, merton1971theory}. 

An option contract is a financial derivative that grants the holder the right, but not the obligation, to buy (call) or sell (put) a specific security at a predetermined price. European-style options can only be exercised on their expiration date, whereas American-style options can be exercised at any time before or on the expiration date. To determine fair option prices, it is essential to understand how underlying asset prices evolve over time. The Black-Scholes (BS) equation is a cornerstone in providing a theoretical framework for it. The BS equation is based on the idea that stock prices follow a random walk and can be modeled using a Gaussian distribution \cite{king1966market, lo2011non}. By assuming that price movements are continuous the Black-Scholes equation is given by:

\begin{equation}
\frac{\partial V}{\partial t} + \frac{1}{2} \sigma^2 S^2 \frac{\partial^2 V}{\partial S^2} + rS \frac{\partial V}{\partial S} - rV = 0    
\end{equation}

where:
\begin{itemize}
    \item \( V(S,t) \) is the price of the derivative (e.g., an option) as a function of the underlying asset price \( S \) and time \( t \).
    \item \( \sigma \) is the volatility of the underlying asset.
    \item \( r \) is the risk-free interest rate.
    \item \( S \) is the price of the underlying asset.
    \item \( t \) represents time.
\end{itemize}
The boundary condition at maturity time \( T \) is:
\begin{equation}
V(T, S) = K(S), \quad \forall S
\end{equation}
where \( K(S) \) is the payoff function of the option. For example, for a European call option:
\begin{equation}
    K(S) = \max(S - K, 0)
\end{equation}
where \( K \) is the strike price.
For a European call option, the closed-form solution is given by the famous \textit{Black-Scholes formula}:
\begin{equation}
    C(S,t) = S N(d_1) - Ke^{-r(T-t)} N(d_2)
\end{equation}
where:
\[
d_1 = \frac{\ln(S/K) + (r + \frac{1}{2} \sigma^2)(T-t)}{\sigma \sqrt{T-t}}, \quad d_2 = d_1 - \sigma \sqrt{T-t}
\]
and \( N(\cdot) \) is the cumulative distribution function of the standard normal distribution.

% statement on limitations of BS 

% Analytical and numeircal methods in the lit to solve BS

This study originally aimed to solve a slightly modified version of the Black-Scholes equation for European options with constant dividends using deep neural networks. However, this specific problem is relatively straightforward and lacks practical significance. A more meaningful challenge lies in pricing American-style options, which do not have closed-form solutions and require numerical methods such as finite differences \cite{brennan1977valuation}. While deep learning approaches have been explored for this problem \cite{han2018deep, becker2019deep, stable_multilevel2022}, their implementation requires a deeper understanding of computational finance techniques, which was beyond the initial scope of this study.

This realization led us to reframe our research question: \emph{Can a deep learning model, developed with limited resources and by a researcher new to machine learning, achieve or even exceed the pricing accuracy of the Nobel Prize-winning Black-Scholes formula?} To explore this, we trained a deep neural network on Petrobras option contracts and evaluated its performance against the Black-Scholes model in predicting call option prices.

Neural networks have been applied to option pricing since the early 1990s \cite{hutchinson1994nonparametric, malliaris1993neural}. Malliaris and Salchenberger \cite{malliaris1993neural} used a single-hidden-layer model to estimate closing prices for S\&P 100 options. Their approach demonstrated comparable accuracy to the Black-Scholes model in approximately half of the test periods. Subsequent research \cite{garcia2000pricing} built upon their work, employing deeper architectures and ensemble methods to improve generalization. For instance, recent studies have explored multi-layer perceptrons (MLPs) with up to four hidden layers and hundreds of neurons per layer, significantly enhancing predictive performance \citep{gan2024option}. Additionally, ensemble techniques have been employed \cite{wang2022option, ruf2019neural} to combine outputs from multiple independent MLPs, further improving robustness.

More recently, deep learning advancements have enabled substantial improvements in option pricing speed and accuracy. Studies such as Zhou \& Wang\cite{multi_asset2023} and Chen \& Liu \cite{accelerated_american2023} highlight the ability of deep neural networks (DNNs) to significantly accelerate calculations while maintaining high precision. However, our primary goal is not to enhance speed but to assess whether using limited computational resources, a deep learning model can still produce reliable pricing predictions. To test this, we developed a simple residual network  trained the model on a consumer-grade RTX 4060 laptop GPU, demonstrating that competitive results can be achieved even without access to high-performance computing infrastructure.

Methodologically, our approach differs from studies such as \cite{deSouzaSantos2024}, which attempted to solve the Black-Scholes partial differential equation. Instead, we focus on direct price prediction employing input feature selection based on $S$, $K$, time to expiration, $\sigma$, and interest rate (as network inputs), following best practices from the approach of Ke and Yang \cite{CS230_2019}. Furthermore, this study builds on the work of Gan and Liu \cite{gan2024option} demonstrating the effectiveness of residual neural networks in financial modeling.

A particularly relevant precedent is the application of deep learning to option pricing in emerging markets. For example, Yilmaz \& Kaya \cite{turkish_emerging2021} explored neural networks in the Turkish market, where higher volatility and lower liquidity create challenges distinct from those in developed markets. Inspired by this work, our study applies a similar approach to the Brazilian market, focusing on Petrobras options, one of the most actively traded contracts in Brazil.

The central research questions of this study are:
\begin{enumerate}
    \item \textit{Can a deep learning model predict Petrobras option prices as accurately as or better than the Black-Scholes model?}
    \item \textit{How does the performance of the neural network vary with different expiration times and price ranges?}
    \item \textit{What are the limitations of using machine learning in this financial modeling context?}
\end{enumerate}

To answer these questions, firstly a dataset was constructed consisting of Petrobras option contracts, their underlying stock prices, interest rates (SELIC), and key financial indicators. We then developed a deep neural network and a custom loss function trained to estimate call option prices based on these features. The performance of the model was compared to the Black-Scholes pricing formula, evaluating accuracy based on real market data. 

\section{Data Collection and Processing}
Collecting data for Petrobras option contracts was more challenging than expected. Unlike American market data, which is readily available via the yfinance API, this data was not conveniently compiled.
Several providers offer access to B3 data, so the easiest solution would have been to purchase the data from one of these services, but due to a lack of funding, this was not an option.
For the present study, four things were needed: The option data, the price of Petrobras stock, the dividends of the Petrobras stock, and the Brazilian interest rate (SELIC).
For the SELIC rate, the data was taken directly from the \href{https://www.bcb.gov.br/en/monetarypolicy/selicrate}{official website of the Central Bank of Brazil}:

\begin{center}
    \texttt{https://www.bcb.gov.br/en/monetarypolicy/selicrate},
\end{center}

and for the stock price of Petrobras, an aggregator was used.
For the option data, the official B3 website was used.\textsuperscript{\dag}\footnotetext{The Black-Scholes equation was applied without incorporating dividends, as their inclusion tends to render analytical price estimations less representative of actual Petrobras option prices. Consequently, obtaining Petrobras' dividend rate was deemed unnecessary. Given that the Black-Scholes model generally underestimates option prices, the addition of the dividend yield term $q$, which acts as a subtractive factor, would further decrease the calculated option prices, reinforcing this tendency.
} To collect the training data from the web sources, two options were considered. \\

\noindent \textbf{Web Scraping with Selenium:}
B3 only provides daily historical data, meaning that if taking this data manually, it would be needed to navigate through each trading day, selecting a box, downloading, and then changing the date on the calendar. Given the large time period needed, this process would have taken an unreasonable amount of time.

To automate this process, \textit{Selenium}, a Python library that allows for automated web interactions by controlling a web browser through scripts, was used. It was able to replicate the manual process of navigating the B3 website, selecting the necessary options, and downloading the files one by one.
However, Google Chrome flagged these automated downloads as potential security threats, requiring it to manually approve each download attempt. While this approach was significantly more manageable than retrieving the data manually, it was still not as efficient as hoped.
\href{https://drive.google.com/file/d/1EKebc71tb8zgyZp5kXGj-qG-v1LxzV3h/view?usp=sharing}{Video of web scraping} \\

\noindent \textbf{Direct HTTP requests:}
The solution was instead of using web scraping, to use the direct download URL pattern used by B3:
\begin{center}
    \texttt{https://www.b3.com.br/pesquisapregao/download?filelist=PEyymmdd.ex\_}
\end{center}
where \texttt{yymmdd} represents the date.
Using Python's \texttt{requests} library, the download process was automated by 
generating URLs for a given date range, sending HTTP requests to retrieve each file, checking the response validity, and saving the files locally.
To accelerate the process, a \textit{ThreadPoolExecutor} was used to run up to 30 concurrent download threads, significantly reducing the total time to get all the data.
This method successfully retrieved the full dataset without manual intervention, making it less time consuming to get the required data. 
\subsection{Data Selection}
To train, validate, and test the model, approximately \textit{eight years of market data} (from \textit{November 2016 to January 2025}) were used. The selection of this timeframe was not based on a strictly scientific approach but rather on the intuition that it would be sufficient to capture market trends and fluctuations in Brazilian interest rates.
It was also needed to define the time to expiration for the selected option contract. In general, shorter-term options tend to be more liquid. Additionally, the Black-Scholes model is known to perform better for short-term options, as its assumptions become less accurate for long expiration periods due to factors such as volatility changes and market frictions, so options expiring in 1, 2, and 3 months were used.
\subsection{Accessing and Cleaning up the Data}
B3 provides daily stock market data in compressed files, but they are not immediately usable in their raw format. Processing them required multiple steps, from extracting the files to filtering only the relevant option contracts. However, B3 provides a structured explanation of the data format for options, which helped streamline the process. The key details are presented in table~\ref{tab:b3_format}:
\begin{table}[h!]
    \caption{B3 File Format Specification}
    \centering
    \begin{tabular}{lll}
        \toprule
        \textbf{Field} & \textbf{Format} & \textbf{Description} \\
        \midrule
        \multicolumn{3}{l}{\textbf{HEADER}} \\
        File generation date & Alphanumeric (A) & Date the file was created (YYYYMMDD) \\
        \midrule
        \multicolumn{3}{l}{\textbf{BODY}} \\
        Option code & Alphanumeric (A) & Unique identifier for the option \\
        Separator & Alphanumeric (A) & Fixed value: “,” (comma) \\
        Option type & Alphanumeric (A) & “C” for call options, “V” for put options \\
        Separator & Alphanumeric (A) & Fixed value: “,” (comma) \\
        Option style & Alphanumeric (A) & “A” for American options, “E” for European options \\
        Separator & Alphanumeric (A) & Fixed value: “,” (comma) \\
        Expiration date & Numeric (N) & Expiration date of the option (YYYYMMDD) \\
        Separator & Alphanumeric (A) & Fixed value: “,” (comma) \\
        Strike price & Numeric (N) & Exercise price (decimal separator: ".") \\
        Separator & Alphanumeric (A) & Fixed value: “,” (comma) \\
        Reference premium & Numeric (N) & Option premium (decimal separator: ".") \\
        Separator & Alphanumeric (A) & Fixed value: “,” (comma) \\
        Implied volatility & Numeric (N) & Volatility per strike price (decimal separator: ".") \\
        \bottomrule
    \end{tabular}
    \label{tab:b3_format}
\end{table}
The downloaded files were in the \texttt{.ex} format, which consists of compressed archives. As these files could not be opened directly, an automated process was implemented to rename them to the \texttt{.zip} extension and extract their contents using Python.
After extraction, the dataset comprised multiple text files distributed across various directories. To facilitate access and organization, all \texttt{.txt} files were put into a dedicated folder, with each file prefixed by the corresponding trading date extracted from the folder name.
The dataset contained information on all listed options, but the focus of this analysis was limited to Petrobras (\texttt{PETR}) European-style contracts. To isolate these contracts, each text file was scanned, selecting only the rows where the first field contained \texttt{PETR} and the third field indicated a European contract (\texttt{E}). Additionally, expired options were excluded by comparing their expiration date with the trading date, retaining only contracts set to expire within one to three months.

\subsection{Data Combination for Training}

To create a single dataset suitable for model training, we needed to combine the Petrobras option data, the PETR4 stock price (including its volatility), and the SELIC rate. First, we computed the volatility of PETR4 by taking the daily percentage changes in the stock price and applying a 21-day rolling standard deviation. This rolling standard deviation was then multiplied by \(\sqrt{252}\) (an approximate number of trading days in a year) to convert it into an annualized figure. 
Next, we brought together all the options data, discarded any contracts that had already expired on or before the trading date, and concatenated all the remaining records into a single table.

Once the option records were completed, each option was matched with the most recent available PETR4 stock price, PETR4 volatility, and SELIC rate as of the date of the option's trade. If an exact date match was missing for any of these fields, we used the closest preceding date instead. This step ensured that all relevant information for each option entry was aligned with the correct trading day.
Finally, we calculated the time-to-expiration (TTE) by subtracting the trade date from the expiration date and converting the result into years. The final dataset, covering eight years of market data, was saved as a single CSV file containing all essential fields: trade date, expiration date, strike price, option premium, PETR4 stock price, PETR4 volatility, SELIC rate, and ticker. This structured dataset presented in Table \ref{tab:csv_sample} was then used for model training and evaluation.

\begin{table}[h!]
    \caption{Sample of the data from the CSV file}
    \centering
    \setlength{\tabcolsep}{3pt} % Adjust column spacing (default is 6pt)
    \begin{tabular}{ccccccccccc}
        \toprule
        trade\_date & ticker  & call/put & exercise & expiration & strike & premium & last\_field & price & return & selic\_rate \\
        \midrule
        2016-11-04 & PETRA34 & C & E & 2017-01-16 & 19.25 & 0.56 & 51.14 & 15.34 & -0.0084 & 14.0 \\
        2016-11-04 & PETRA70 & C & E & 2017-01-16 & 10.75 & 5.60 & 60.85 & 15.34 & -0.0084 & 14.0 \\
        2016-11-04 & PETRA4  & C & E & 2017-01-16 & 14.75 & 2.34 & 51.79 & 15.34 & -0.0084 & 14.0 \\
        \multicolumn{11}{c}{\vdots} \\
        \bottomrule
    \end{tabular}
    \label{tab:csv_sample}
\end{table}

\newpage
\section{Methodology}
\subsection{Machine Learning Model}
As the work of \citep{CS230_2019}, we selected the input features of the deep neural network to align with the variables required to solve the Black-Scholes equation: $\ln(S)$, $\ln(K)$, \texttt{tte}(time to expiry), \texttt{vol}(volatility), and \texttt{selic\_rate}(Brazil's central bank interest rate), and as the output we chose to focus on estimating call option prices, as \cite{CS230_2019} suggests as a possible improvement to their implementation on the American market. 

% \begin{figure}[h!]
%     \centering
%     \includegraphics[width=0.4\linewidth]{Images/deep_neural_network_draft.png}
%     \caption{General Idea of the Deep Neural Network to Predict Call Prices}
%     \label{fig:enter-label}
% \end{figure}

\begin{figure}[h!]
    \centering
    \includegraphics[width=\linewidth]{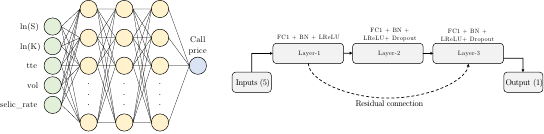}
    \caption{(Left) Deep neural network architecture adapted from \cite{CS230_2019}, where the input features represent key financial variables, and the output corresponds to the call option price. (Right) The residual deep neural network used in this work, featuring a residual connection between Layer 1 and Layer 2. Abbreviations: FC – Fully Connected Layer, BN – Batch Normalization, LReLU – Leaky ReLU Activation.}
    \label{fig:net_overview}
\end{figure}

% \subsubsection{Network Architecture} 

% MENTION THE RESNET nature article 
% https://link.springer.com/article/10.1007/s10614-023-10413-3

Drawing inspiration from computer vision, particularly the use of deep residual networks \citep{He2016}, we devised a ResNet-inspired architecture that is tailored to the option pricing problem. We believed that incorporating residual connections would help capture the non-linear interactions among the input features derived from the Black-Scholes framework \citep{gan2024option}. 

In this architecture, each fully connected layer is augmented with batch normalization \citep{Ioffe2015} to stabilize the learning process and accelerate convergence. We selected LeakyReLU activation functions to mitigate the risk of neuron "death" often encountered with standard ReLU unit, as \cite{CS230_2019} did in their MLP architecture. Furthermore, we incorporated dropout layers after key network blocks to reduce the risk of overfitting.

% \begin{figure}[h]
%     \centering
%     \includegraphics[width=0.9\linewidth]{Images/Network_arch.png}
%     \caption{Selected architecture for training}
%     \label{fig:arch}
% \end{figure}

\subsection{Loss Function and Training}

To align the network’s predictions with both observed market prices and established theoretical models, we developed a custom hybrid loss function. This loss function combines two mean squared error (MSE) components: one that measures the error between the predicted option prices and the market-observed prices, and another that compares the predictions with the Black–Scholes calculated prices. A dynamic weighting scheme—implemented via a masking strategy—allows the model to learn from the empirical data while remaining grounded in theoretical benchmarks.

Let \( p_i \) be the predicted option price, \( m_i \) the market-observed price, and \( bs_i \) the Black–Scholes calculated price for the \( i \)th sample. We first define a binary mask \( \gamma_i \) that indicates whether the market price is considered reliable:

\[
\gamma_i = 
\begin{cases}
1, & \text{if } m_i > 0.1, \\
0, & \text{otherwise.}
\end{cases}
\]

The loss for each sample is given by a weighted combination of two mean squared error (MSE) components:
\[
\ell_i = \gamma_i \, (p_i - m_i)^2 + (1 - \gamma_i) \, (p_i - bs_i)^2.
\]

Averaging over all \(N\) samples, the overall hybrid loss function is:
\[
\mathcal{L} = \frac{1}{N} \sum_{i=1}^{N} \ell_i = \frac{1}{N} \sum_{i=1}^{N} \left[ \gamma_i \, (p_i - m_i)^2 + (1 - \gamma_i) \, (p_i - bs_i)^2 \right].
\]

This dynamic weighting scheme enables the model to prioritize learning from the market data when the observed prices are reliable, while defaulting to the theoretical Black--Scholes prices when the market data is less informative.

For training, we adopted the Adam optimizer \cite{Kingma2015} and conducted an extensive hyperparameter search using the Optuna framework \citep{Akiba2019}. Our search encompassed key parameters such as the learning rate, hidden layer size, dropout probability, and weight decay. Each candidate configuration was evaluated over 10 epochs with the validation loss guiding the selection process. Once the optimal hyperparameters were identified, the final model was retrained for 25 epochs with early stopping to prevent overfitting. Throughout this process, training and validation loss curves were monitored to ensure proper convergence and to diagnose any potential issues in the learning dynamics.

Lastly, we divided the data into three sets: train, validation, and test, to ensure a robust evaluation and comparison of our results. The training and validation sets were drawn from the period between November 2016 and October 2024, with an 80-20 split. The test set comprised data from November, December, and January of 2025.

\section{Results and discussion}
\subsection{Hyperparameter optimization and final training losses}
\begin{figure}[h]
    \centering
    \includegraphics[width=0.7\linewidth]{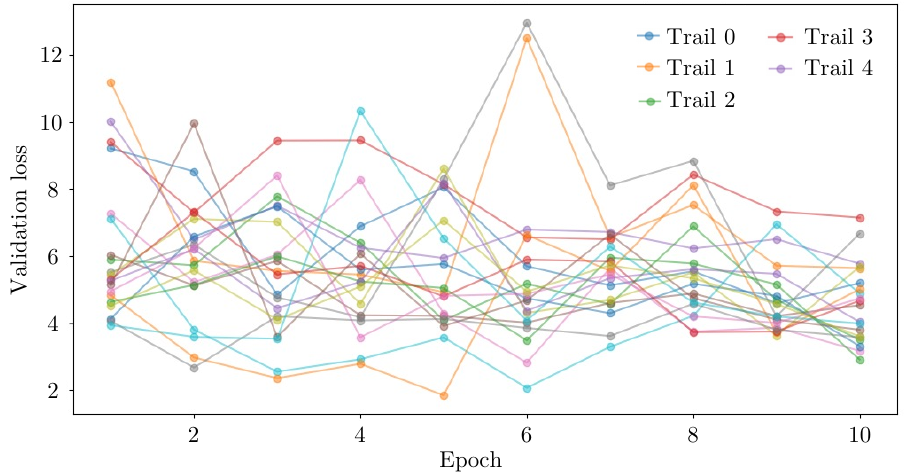}
    \caption{Epoch vs. loss plots for each hyperparameter trail performed using \texttt{optuna} with the validation dataset.}
    \label{fig:hyp_loss}
\end{figure}

Before the final model training, a simple hyperparameter search was performed using the Optuna framework. This process explored key parameters-including the learning rate, hidden layer size, dropout probability, and weight decay to determine the optimal configuration for our model. The plot in Fig. \ref{fig:hyp_loss} illustrates the model's convergence behavior and training stability under different configurations. Although our hyperparameter search produced detailed loss curves that guided our final parameter selection, the plot in Fig. \ref{fig:hyp_loss} is not the original loss graph from that process. Instead, it serves as a representative example, illustrating the model's convergence behavior. Only a subset of the hyperparameter search results is shown here for illustrative purposes.

% Although the hyperparameter search generated detailed loss curves that guided the parameter selection, the plot shown below \ref{fig:hyp_loss} is not the original loss graph from that process. Instead, it is a representative loss curve from a separate training run. Unfortunately, the original graph was lost and, due to time constraints, the model could not be re-train to reproduce it.

The optimal hyperparameters identified were:
\[
\{\texttt{learning rate}: 5.74\times10^{-5},\ \texttt{hidden\_size}: 256,\ \texttt{dropout\_p}: 0.2414,\ \texttt{weight\_decay}: 1.93\times10^{-4}\}.
\]
The notably small weight decay value suggests that its regularization effect may be minimal or even unnecessary for this architecture. This insight warrants further investigation into alternative regularization techniques or even the omission of weight decay to optimize model performance.

\begin{figure}[h]
    \centering
    \includegraphics[width=0.7\linewidth]{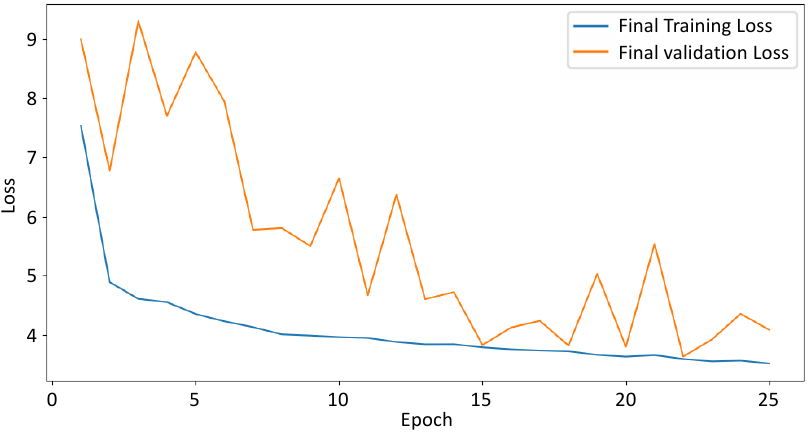}
    \caption{Evolution of training and test loss during the final training.}
    \label{fig:best_loss}
\end{figure}

Fig. \ref{fig:best_loss} presents the loss evolution on both training and validation datasets. The validation loss exhibits an overall decreasing trend, albeit with some fluctuations, alongside the training loss, suggesting a favorable learning process, as shown in Figure \ref{fig:best_loss}.

\subsection{Comparison with the Black-Scholes Solution}
The performance of the model was evaluated using Mean Absolute Error (MAE), defined as:
\[
    \text{MAE} = \frac{1}{n} \sum_{i=1}^{n} |y_i - \hat{y}_i|
\]
where:
\begin{itemize}
    \item $n$ is the total number of observations,
    \item $y_i$ is the actual value for observation $i$,
    \item $\hat{y}_i$ is the predicted value for observation $i$.
\end{itemize}

\begin{figure}[h!]
    \centering
    \begin{subfigure}{0.48\textwidth}
        \centering
        \includegraphics[width=\linewidth]{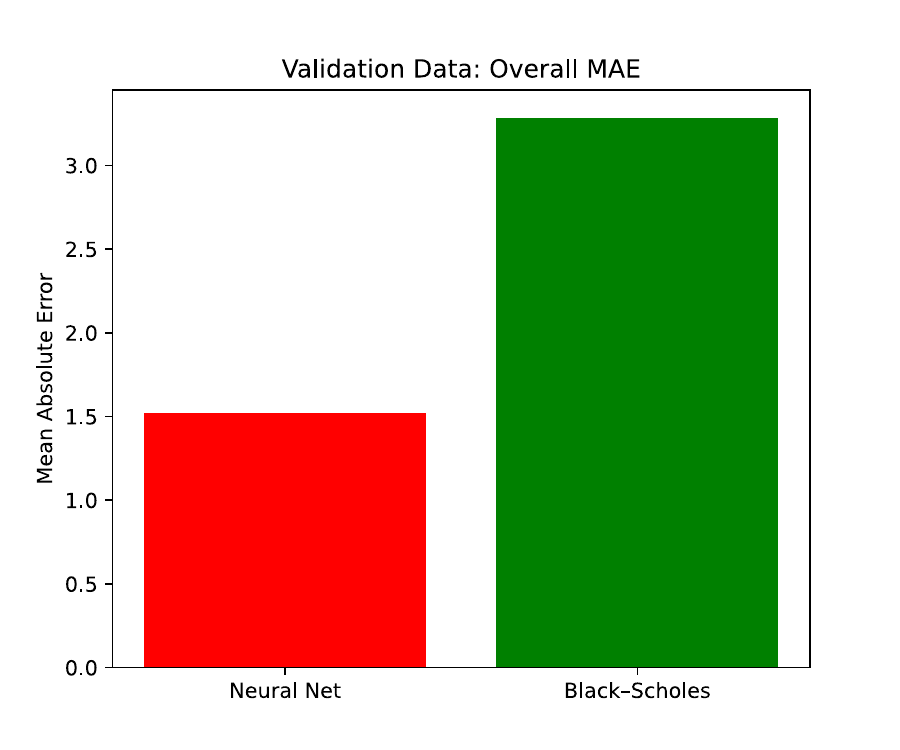}
        \caption{Validation – Overall Error Analysis}
    \end{subfigure}
    \begin{subfigure}{0.48\textwidth}
        \centering
        \includegraphics[width=\linewidth]{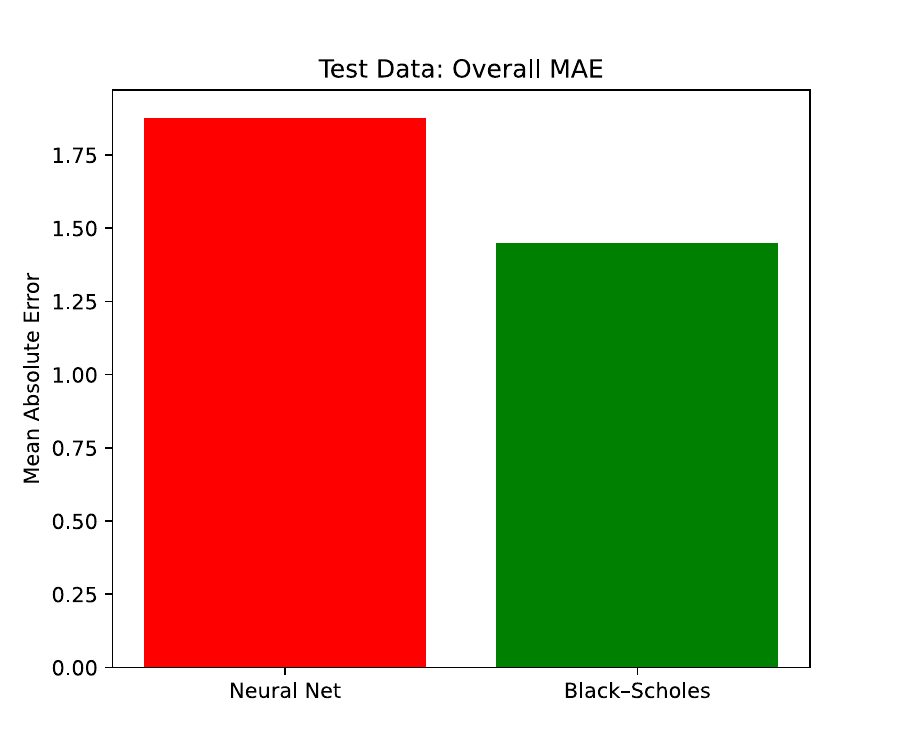}
        \caption{Test – Overall Error Analysis}
    \end{subfigure}
    \caption{Comparison of overall error for validation and test sets}
    \label{fig:error_analysis_overall}
\end{figure}

\begin{figure}[h!]
    \centering
    \includegraphics[width=\linewidth]{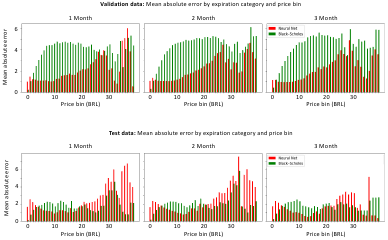}
    \caption{Comparison of mean absolute errors for the validation (top) and test (bottom) sets, based on option expiration time and strike price.}
    \label{fig:mae_monthwise}
\end{figure}

To gain deeper insights, we also analyzed MAE errors by month and for price brackets of 1 BRL (Brazilian real - currency). The machine learning model consistently outperforms or closely matches the Black-Scholes model for prices ranging from 3 to 19 BRL in the test set, as illustrated in Fig. \ref{fig:mae_monthwise}. For this price range, the model achieves a significant improvement of a 64.3\% reduction in MAE across all months for around  43.41\% of all Petrobras option transactions on B3. 

% \begin{figure}[h]
%     \centering
%     \makebox[\textwidth][c]{ % Centers content within text width
%         \begin{subfigure}{1.30\textwidth}
%             \includesvg[width=\linewidth]{Results_Images/error_analysis_validation_by_exp_and_price.svg}
%             \caption{Validation - Error Analysis by Expiration and Price}
%         \end{subfigure}
%     }
    
%     \makebox[\textwidth][c]{ % Centers second subfigure as well
%         \begin{subfigure}{1.30\textwidth}
%             \includesvg[width=\linewidth]{Results_Images/error_analysis_test_by_exp_and_price.svg}
%             \caption{Test - Error Analysis by Expiration and Price}
%         \end{subfigure}
%     }
    
%     \caption{Comparison of Error Analysis by Expiration and Price for Validation and Test Sets}
% \end{figure}

\begin{table}[h!]
    \centering
    \caption{Mean Absolute Error (MAE) by Expiration Category}
    \label{tab:mae_expiration}
    \begin{tabular}{lcc}
        \toprule
        \textbf{Expiration Category} & \textbf{MAE (Deep Learning)} & \textbf{MAE (Black–Scholes)} \\
        \midrule
        1 Month  & 1.2873 & 1.9975 \\
        2 Months & 1.3663 & 2.0596 \\
        3 Months & 1.1932 & 2.0089 \\
        \midrule
        \multicolumn{3}{l}{\textbf{Overall MAE in [3, 19]:}} \\
        \multicolumn{2}{l}{Deep Learning: \textbf{1.3047}} & Black–Scholes: \textbf{2.0295} \\
        \bottomrule
    \end{tabular}
\end{table}

% \begin{figure}[h]
%     \centering
%     \begin{subfigure}{0.48\textwidth}
%         \centering
%         \includesvg[width=\linewidth]{Results_Images/val_scatter.svg}
%         \caption{Validation - Daily Averages}
%     \end{subfigure}
%     \hfill
%     \begin{subfigure}{0.48\textwidth}
%         \centering
%         \includesvg[width=\linewidth]{Results_Images/test_scatter.svg}
%         \caption{Test - Daily Averages}
%     \end{subfigure}
%     \caption{Comparison of Daily Average Prices for Validation and Test Sets}
% \end{figure}

\begin{figure}[h]
    \centering
    \includegraphics[width=\linewidth]{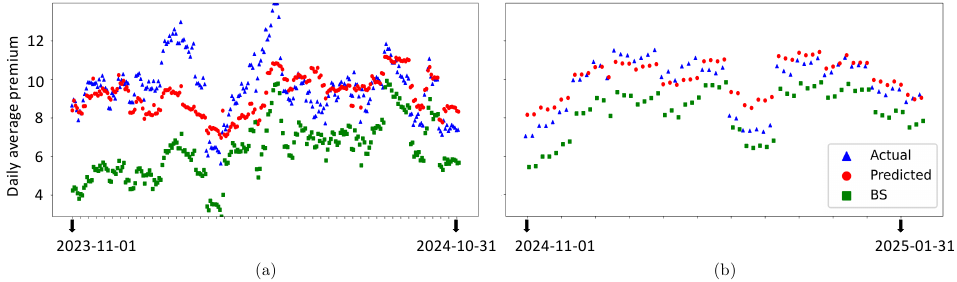}
    \caption{Comparison of daily average prices: ground truth (actual), neural network predictions, and Black-Scholes estimates for the validation set (left) and test set (right).}
    \label{fig:dailyAverage}
\end{figure}

\begin{figure}[h!]
    \centering
    \includegraphics[width=\linewidth]{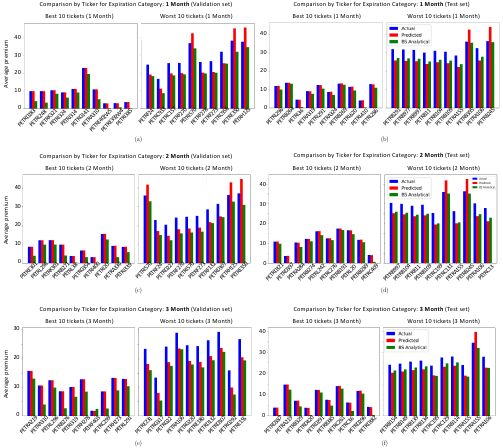}
    \caption{Comparison of validation and test results for 1, 2, and 3 Months expiration category.}
    \label{fig:123Months}
\end{figure}

We also examined how the mean prices predicted by both the Black-Scholes model and the neural network compare to the observed daily mean prices. As shown in Figure \ref{fig:dailyAverage}, the neural network predictions align more closely with the observed values than those of the Black-Scholes model. Additionally, we examined the 10 best- and worst-performing tickers for each month (Figure \ref{fig:123Months}). Notably, for certain assets priced around 5 BRL, such as PETRL36, PETRL410, PETRD397, and PETR409, significant discrepancies were observed in the test set, where the Black-Scholes model consistently undervalued them.

\newpage
\subsection{Discussion}
% This study aimed to assess whether a deep learning model could predict Petrobras option prices as accurately as or better than the Black-Scholes model. 

The results indicate that the deep learning model significantly outperforms the Black-Scholes model within the 3-19 BRL price range, reducing the mean absolute error (MAE) by 64.3\% across all months for 43.41\% of Petrobras option transactions on B3. Additionally, it successfully values contracts priced around 5 BRL, such as PETRL36, PETRL410, PETRD397, and PETR409, whereas the Black-Scholes model severely undervalues them. However, the deep learning model also has limitations, particularly a tendency to overestimate option prices outside the 3-19 BRL range.

In terms of performance variation across different expiration times and price ranges, the neural network demonstrates a counterintuitive trend: its accuracy improves as it is trained on options with longer expiration periods, in contrast to Black-Scholes and conventional financial intuition, which typically expect shorter-term options to be easier to price. Furthermore, qualitative analysis of the model’s outputs reveals that its daily average predictions align more closely with observed market prices.

Regarding the limitations of using machine learning in this financial modeling context, one key challenge is the model's weaker generalization to extreme pricing scenarios, where deviations from market behavior become more pronounced. Additionally, while data-driven approaches like deep learning can capture complex patterns, they remain sensitive to training data biases, which may hinder their robustness in unseen market conditions. Despite these challenges, the results suggest that deep learning presents a viable alternative to traditional models for specific price ranges and contract structures, warranting further exploration in financial modeling.
A possible improvement is to have different models for different price ranges and expiration dates.

\section{Summary}

This study investigated the application of deep residual networks for option pricing, specifically focusing on Petrobras option contracts in the Brazilian market. Traditional models, such as the Black-Scholes equation, provide a theoretical framework for pricing financial derivatives but often fail to capture real market dynamics accurately. This research aimed to determine whether a deep learning model trained with minimal resources can predict option prices more accurately than the Black-Scholes model.

To achieve this, historical Petrobras option data was collected from B3, the Brazilian stock exchange, along with stock prices and interest rates. Due to the lack of publicly available, consolidated data sources, a combination of web scraping, direct HTTP requests, and automated data processing techniques was employed to compile a comprehensive dataset spanning from November 2016 to January 2025.

A deep neural network was designed, inspired by residual network architectures commonly used in computer vision. The model was trained using a custom hybrid loss function that dynamically weights empirical market data and theoretical Black-Scholes prices. Hyperparameter tuning was performed using Optuna, leading to an optimized architecture that included batch normalization, dropout layers, and LeakyReLU activations.

The model's performance was evaluated using mean absolute error (MAE) across different price ranges and expiration periods. Results show that the deep learning model consistently outperforms Black-Scholes in predicting option prices within the 3-19 BRL range, achieving a 64.3\% reduction in MAE for 43.41\% of all Petrobras option transactions on B3. Additionally, the model successfully identified cases where the Black-Scholes model significantly undervalued options, such as PETRL36, PETRL410, PETRD397, and PETR409.

Despite its success in specific price ranges, the neural network exhibited limitations in predicting prices outside the 3-19 BRL range, tending to overestimate prices in these cases. Additionally, while the model showed improved accuracy for longer expiration periods—contrary to conventional financial expectations—its generalization to extreme pricing scenarios remains an open challenge. Furthermore, signs of overfitting were observed which restricted us to run only up to 25 epochs. This will be addressed in future work as we attempt to increase the predicted call price accuracy by employing more robust models.

The findings suggest that deep learning may provide a promising alternative to traditional option pricing models for specific market conditions. Future work could explore using separate models for different price ranges and expiration periods to improve accuracy further. % The study underscores the potential of machine learning in financial modeling and highlights areas for further refinement to enhance robustness and predictive performance.

\bibliographystyle{unsrt} 
\bibliography{Deep_Residual_Networks_for_Option_Pricing} 

\end{document}